\documentstyle[twocolumn,aps,epsfig]{revtex}
\begin{document}
\draft
\title{Antideuteron yield at the AGS and coalescence implications}
%\begin{center}
\bigskip
%\end{center}
\author{
T. A. Armstrong               \unskip,$^{(8,\ast)}$
K. N. Barish                  \unskip,$^{(3)}$
S. Batsouli                   \unskip,$^{(13)}$
S. J. Bennett                 \unskip,$^{(12)}$
M. Bertaina                   \unskip,$^{(7,\dag)}$
A. Chikanian                  \unskip,$^{(13)}$\\
S. D. Coe                     \unskip,$^{(13,\ddag)}$
T. M. Cormier                 \unskip,$^{(12)}$
R. Davies                     \unskip,$^{(9,\S)}$
C. B. Dover                   \unskip,$^{(1,\|)}$
P. Fachini                    \unskip,$^{(12)}$
B. Fadem                      \unskip,$^{(5)}$
L. E. Finch                   \unskip,$^{(13)}$\\
N. K. George                  \unskip,$^{(13)}$
S. V. Greene                  \unskip,$^{(11)}$
P. Haridas                    \unskip,$^{(7,\P)}$
J. C. Hill                    \unskip,$^{(5)}$
A. S. Hirsch                  \unskip,$^{(9)}$
R. Hoversten                  \unskip,$^{(5)}$
H. Z. Huang                   \unskip,$^{(2)}$\\
H. Jaradat                    \unskip,$^{(12)}$
B. S. Kumar                   \unskip,$^{(13,\ast\ast)}$
T. Lainis	              \unskip,$^{(10)}$
J. G. Lajoie                  \unskip,$^{(5)}$
Q. Li                         \unskip,$^{(12)}$
B. Libby                      \unskip,$^{(5,\dag\dag)}$
R. D. Majka                   \unskip,$^{(13)}$\\
T. E. Miller                  \unskip,$^{(11)}$
M. G. Munhoz                  \unskip,$^{(12)}$
J. L. Nagle                   \unskip,$^{(4)}$
I. A. Pless                   \unskip,$^{(7)}$
J. K. Pope                    \unskip,$^{(13,\ddag\ddag)}$
N. T. Porile                  \unskip,$^{(9)}$\\
C. A. Pruneau                 \unskip,$^{(12)}$
M. S. Z. Rabin                \unskip,$^{(6)}$
J. D. Reid                    \unskip,$^{(11)}$
A. Rimai                      \unskip,$^{(9,\S\S)}$
A. Rose                       \unskip,$^{(11)}$
F. S. Rotondo                 \unskip,$^{(13,\|\|)}$\\
J. Sandweiss                  \unskip,$^{(13)}$
R. P. Scharenberg             \unskip,$^{(9)}$
A. J. Slaughter               \unskip,$^{(13)}$
G. A. Smith                   \unskip,$^{(8)}$
M. L. Tincknell               \unskip,$^{(9,\P\P)}$\\
W. S. Toothacker              \unskip,$^{(8)}$
G. Van Buren                  \unskip,$^{(7,2)}$
F. K. Wohn                    \unskip,$^{(5)}$
Z. Xu                         \unskip$^{(13)}$
}
\address{\centerline{(E864 Collaboration)}}
%\vskip \baselineskip
\address{  $^{(1)}$ Brookhaven National Laboratory, Upton, 
New York 11973 \break
  $^{(2)}$ University of California at Los Angeles, Los Angeles, 
California 90095 \break  
  $^{(3)}$ University of California at Riverside, Riverside, 
California 92521 \break
  $^{(4)}$ Columbia University, Nevis Laboratory, Irvington,
New York 10533 \break 
  $^{(5)}$ Iowa State University, Ames, Iowa 50011 \break 
  $^{(6)}$ University of Massachusetts, Amherst, Massachusetts 01003 \break 
  $^{(7)}$ Massachusetts Institute of Technology, Cambridge, 
Massachusetts 02139 \break 
  $^{(8)}$ Pennsylvania State University, University Park, 
Pennsylvania 16802 \break 
  $^{(9)}$ Purdue University, West Lafayette, Indiana 47907 \break 
  $^{(10)}$ United States Military Academy, West Point 10096\break
  $^{(11)}$ Vanderbilt University, Nashville, Tennessee 37235 \break 
  $^{(12)}$ Wayne State University, Detroit, Michigan 48201 \break 
  $^{(13)}$ Yale University, New Haven, Connecticut 06520 \break
}

\date{\today}
\maketitle
%\begin{center}
\begin{abstract}
We present Experiment 864's measurement of invariant
antideuteron yields in 11.5$A$~GeV/c Au~+~Pt collisions.
The analysis includes 250 million triggers representing
14 billion 10\% central interactions sampled for events with high
mass candidates. We find
(1/2$\pi$$p_t$)$d$$^2$$N$/$d$$y$$d$$p_t$ = 
$3.5 \pm 1.5 (stat.) ^{+0.9} _{-0.5} (sys.) \times 10^{-8}$ GeV$^{-2}c^{2}$
for 1.8$<$$y$$<$2.2, $<$$p_t$$>=$0.35~GeV/c ($y_{cm}$=1.6) and
$3.7 \pm 2.7 (stat.) ^{+1.4} _{-1.5} (sys.) \times 10^{-8}$ GeV$^{-2}c^{2}$
for 1.4$<$$y$$<$1.8, $<$$p_t$$>=$0.26~GeV/c,
and a coalescence parameter $\overline{B_2}$ of
$4.1 \pm 2.9 (stat.) ^{+2.3} _{-2.4} (sys.) \times 10^{-3}$ GeV$^{2}c^{-3}$.
Implications for coalescence and
antimatter annihilation are discussed.
 
\end{abstract}
%\end{center}
\pacs{PACS numbers: 25.75.Dw, 25.75.Gz}
\section{Introduction}
The production of antinucleons and antinuclei
in heavy ion collisions is of significant interest
for several reasons~\cite{bmeter,mfld,ssize}.
Because the initial colliding system contains no antibaryons, their yields
and spectra are determined solely by collision dynamics.
At AGS energies ($\sqrt{s}$ = 4.8~A~GeV), nucleon-antinucleon
pair production is above threshold in individual $Nucleon + Nucleon$
collisions, but direct deuteron-antideuteron pair production is not.
However, a small fraction of the colliding nucleon pairs which have
high relative Fermi momenta may exceed the threshold for $d$-$\overline{d}$
pair production.  We have
estimated this contributes at a level at least two orders
of magnitude below the measured yield from our experiment.
Antideuterons are therefore predominantly formed in secondary interactions
between directly produced antinucleons from the collision.
Their production is then highly dependent on the total abundances
and spatial distribution of the antinucleons.

Simple coalescence and thermal models~\cite{zu,bo,sa} indicate that differences
in the measurement of the coalescence parameter between nuclei ($B_A$)
and their antinuclei counterparts ($\overline{B_A}$)
are due to differences in the source volumes, where
\begin{eqnarray}
B_A &\equiv&
\frac{\left( \frac{1}{2 \pi p_t} \frac{d^2N_A}{dydp_t}\right)}
     {\left( \frac{1}{2 \pi p_t} \frac{d^2N_p}{dydp_t}\right)^Z
      \left( \frac{1}{2 \pi p_t} \frac{d^2N_n}{dydp_t}\right)^{A-Z}}\\
&\approx&
\frac{\left( \frac{1}{2 \pi p_t} \frac{d^2N_A}{dydp_t}\right)}
     {\left( \frac{1}{2 \pi p_t} \frac{d^2N_p}{dydp_t}\right)^A},\label{ba}\\
\overline{B_A} &\approx&
\frac{\left( \frac{1}{2 \pi p_t} \frac{d^2N_{\overline{A}}}{dydp_t}\right)}
     {\left( \frac{1}{2 \pi p_t} \frac{d^2N_{\overline{p}}}{dydp_t}\right)
     ^A}. \label{eq}
\end{eqnarray}
We have used an assumption in Equation~\ref{ba} that $n$ and
$p$ abundances in the region of the measurement are similar,
which is only approximately true~\cite{neuts}. Antineutron
yields remain unmeasured.
    
The small binding energy of (anti)deuterons requires that they be formed
near the hypersurface of the fireball where their mean free
path is sufficiently large that they suffer no further collisions.
The additional annihilation cross section for
antinucleons implies that densities must be even lower
for $\overline{d}$s than $d$s
to have sufficient mean free paths for survival~\cite{heinz}, possibly
resulting in a more shell-like spatial formation zone for
$\overline{d}$s~\cite{sm}. This has fueled
predictions that $\overline{B_2}$ will be notably smaller than
$B_2$~\cite{heinz,sm,surf}.

However, other thermal model calculations suggest $B_2$ and $\overline{B_2}$
may be very similar. Microscopic models demonstrate
that nucleons may undergo 20-30 collisions before their final interactions
in heavy ion collisions at AGS and SPS energies~\cite{qgsm,urqmd}.
The more collisions the constituents undergo,
the further towards thermal and chemical equilibrium the system is driven. If
equilibrium is complete,
and all particles freeze-out along the same hypersurface,
no difference would be expected
between $B_2$ and $\overline{B_2}$~\cite{heinz2}. An analysis
of equilibrium conditions in the collisions studied by E864 is
presented elsewhere~\cite{ismd99}.

A sample of two $\overline{d}$s was previously measured in the AGS
Experiment 858 in minimum bias
Si~+~Au collisions at 14.6~GeV/c per nucleon, indicating that in such a system
$\overline{B_2}$ is below or at the level of $B_2$~\cite{ps}.
However, this measurement
involves a smaller system size where antimatter annihilation may not be
as prominent as in central Au~+~Pt collisions. Antiproton yields are 
significantly suppressed from first collision scaling in the larger
colliding systems compared to peripheral collisions or Si + Au
collisions~\cite{db}.
Some transport models~\cite{ssize}~indicate that over 90\% of
the originally produced $\overline{p}$s are annihilated, while other
calculations include a screening of the annihilation in this dense
environment thus reducing the losses~\cite{kahana}.
These large collisions, consequently, provide a good place to see
the effects of annihilation on $\overline{d}$ production.
Interestingly, there may be additional processes in these collisions
whose contributions to antimatter
coalescence rates are not completely understood, such
as findings that a significant portion of the antibaryon number is
carried away in the form of strange antibaryons~\cite{jl}.

\section{EXPERIMENT}

In order to search for rare products from heavy ion collisions at the AGS,
Experiment 864 was designed as an open geometry,
high rate spectrometer~\cite{e864}. It features a
multiplicity detector for triggering on central collisions~\cite{beam}, and a
level 2 trigger capable of selecting events in which high mass objects
traverse the spectrometer, using a hadronic calorimeter to provide
fast energy and time measurements~\cite{let,calo}.
The calorimeter energy measurement encompasses the kinetic energy for normal
hadronic matter, with an additional annihilation energy of approximately
twice the particle mass for antimatter. 
Tracking of charged particles is performed primarily with
three scintillating hodoscope TOF walls
interspersed with two straw tube tracking stations
for improved track spatial resolution. The tracking system is used
to determine velocities and rigidities of charged particles accurately,
providing mass resolutions typically better than 5\%~\cite{e864}.

\begin{figure}
\epsffile{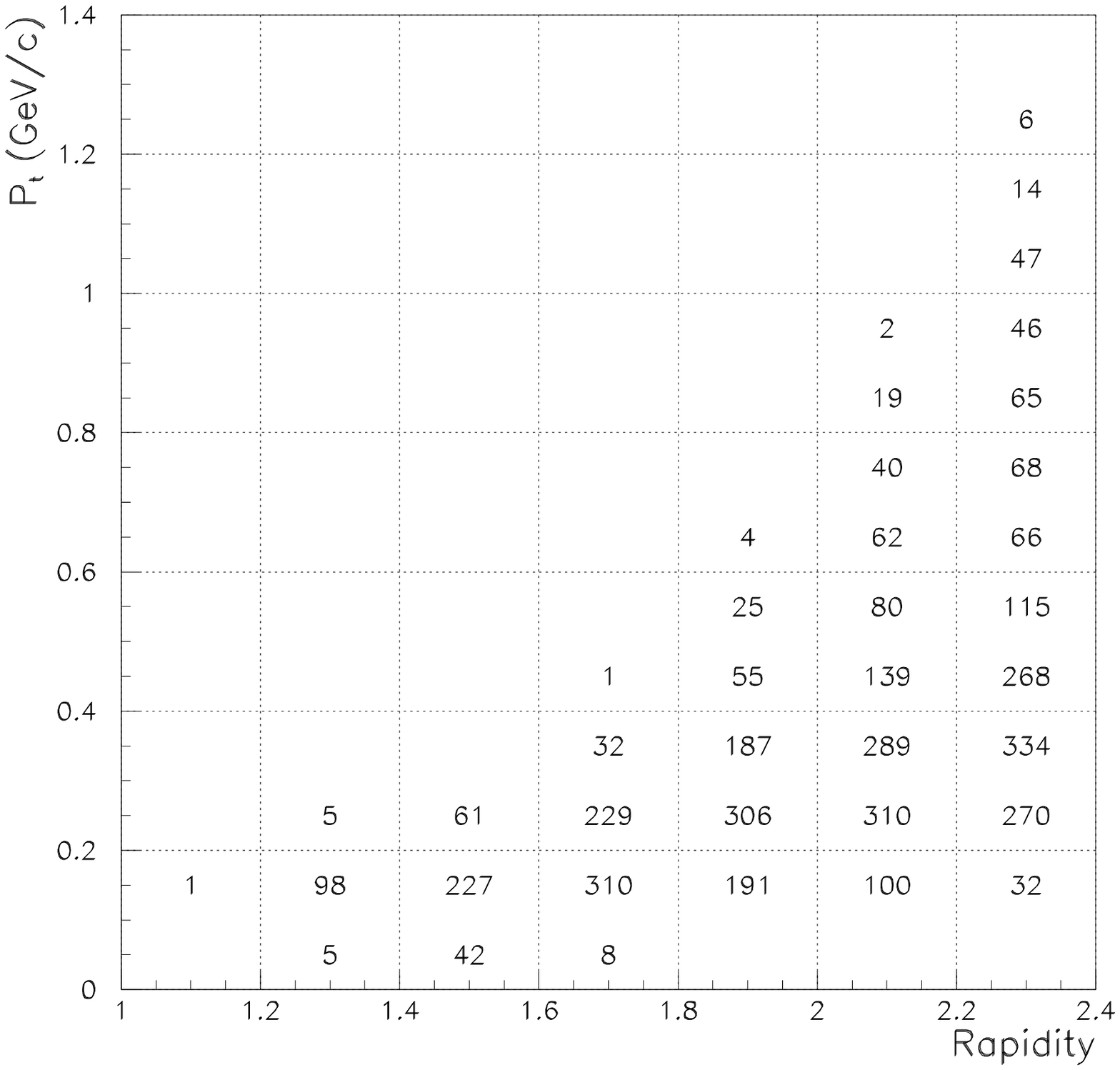}
\caption{Fractional geometric acceptance for antideuterons ($\times$1000) at the
-0.75T field setting for E864 ($y_{cm} = 1.6$).}
\label{fi:accep}
\end{figure}

The spectrometer magnets can be set to different
field strengths and polarities which optimize acceptance for particles of
interest. During the 1996-1997 run of the experiment, over 250 million
triggers were taken with both magnets set to -0.75T. This setting allows
reasonable acceptance for $\overline{d}$s while most positively charged particles
and lighter negatively charged particles are swept out of the acceptance.
Figure~\ref{fi:accep} shows the $\overline{d}$ geometric
acceptance for this field setting.

\section{ANALYSIS}

\begin{figure}
\epsffile{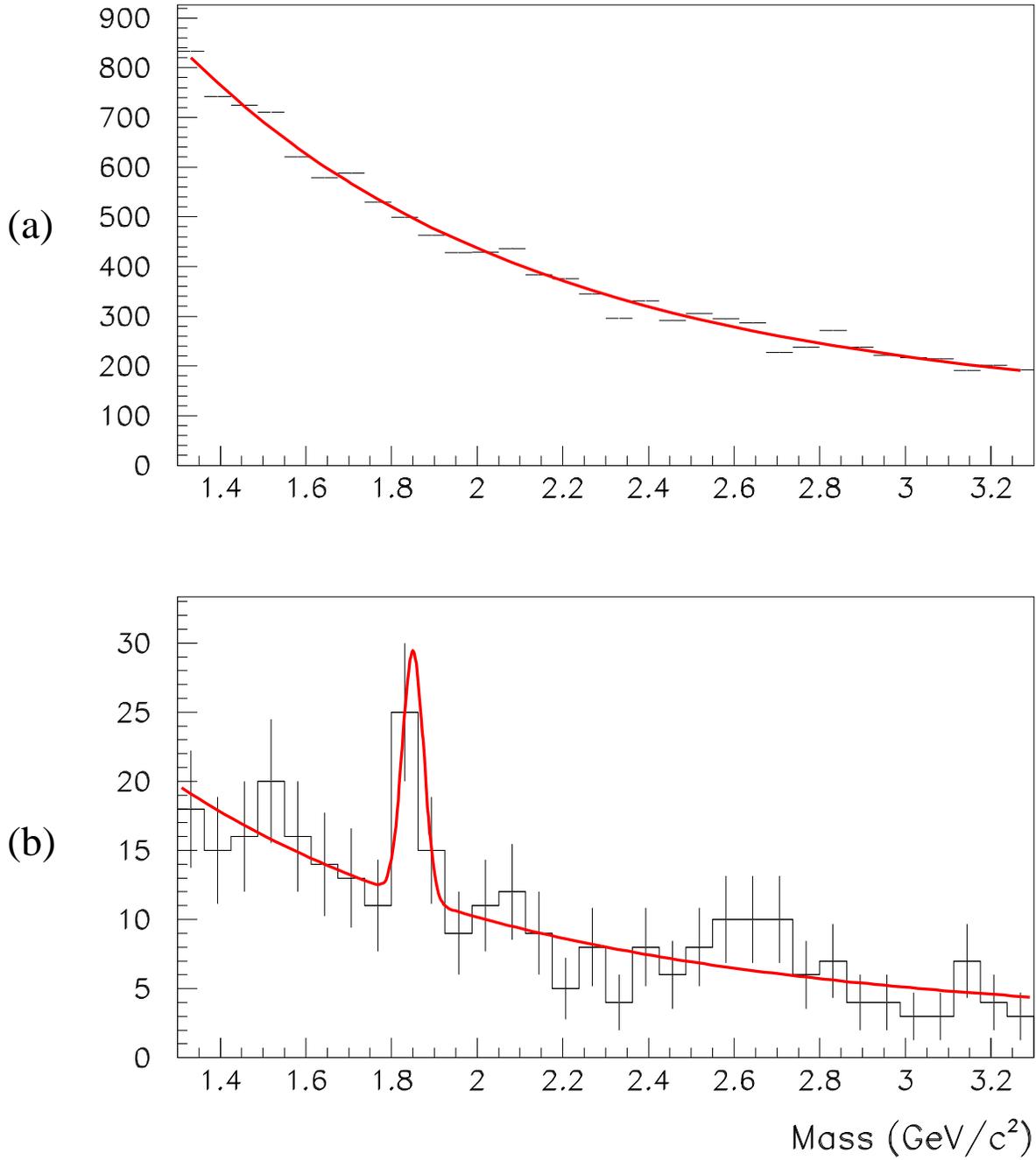}
\caption{Mass spectrum of charge $Z = -1$ particles ($1.8 < y < 2.2$)
before (a) and after (b) a cut is
placed on the energy deposited in the calorimeter to reduce background.}
\label{fi:mass}
\end{figure}

Figure~\ref{fi:mass}a
shows the mass spectrum of charge $-1$ particles from
the data set in the region near the $\overline{d}$ mass ($1.8 < y < 2.2$)
for tracks which have passed basic quality cuts.
Simulations have been performed to study the background evident in the
data, which is understood to have two predominant sources.
In the rapidity range shown in Figure~\ref{fi:mass}a, the background
is dominated by neutrons passing
through the spectrometer magnets before undergoing charge-exchange
reactions ($n + X \rightarrow p + X'$), resulting in stiff tracks which
are calculated with incorrect masses and charge signs.
However, the energy measured
by E864's hadronic calorimeter should be consistent with the kinetic
energy of a $p$ traveling at the speed calculated from the
hodoscope TOF measurements~\cite{calo}. This energy
measurement can then be used to cut away background whose calorimeter
response is reasonably consistent with that of a $p$.
Such a cut is used in the mass spectrum shown in
Figure~\ref{fi:mass}b, revealing a clear peak at the $\overline{d}$
mass. The chosen cut requires that the calorimeter response is greater
than $4\sigma$ above the expected response for a $p$ in order to
remove as much background as possible without destroying the $\overline{d}$
signal (the cut is 95\% efficient for $\overline{d}$s, aided by the significant
energy contributions from annihilation in the calorimeter).
While a $4\sigma$ cut
would normally be expected to remove nearly 100\% of the scattered $p$s,
the efficiency is reduced in this case because of the bias from the
level 2 trigger in selecting candidates whose calorimeter responses are
large. Because this cut is
independent of the background's calculated mass, the shape of the background
is unaffected. A fit to the shape is made before the cut and used in
fitting for signal plus background after the cut. The resulting $\overline{d}$
yield is $17.6 \pm 7.5$ counts, where the statistical error includes
contributions from a background of
$34.1 \pm 5.8 (Poisson~stat.) \pm 2.3 (normalization)$.

Similarly, in the rapidity range between 1.4 and 1.8 the background is
understood to be dominated by $\overline{p}$s which have scattered
by small angles near the spectrometer magnets. Again, the calorimeter
can be used to reject $\overline{p}$s in the mass spectrum, although the
additional energy deposited from annihilation in the calorimeter
leads to a smaller separation from $\overline{d}$s for $\overline{p}$s than
$p$s. A $4\sigma$ cut would severely impinge upon the $\overline{d}$
signal, so a less stringent $2\sigma$ cut is chosen with the
resulting efficiency for $\overline{d}$s at 86\%. A $\overline{d}$
signal is found with $4.6 \pm 3.3$ counts, where the statistical error
includes contributions from a background of
$5.4 \pm 2.3 (Poisson~stat.) \pm 0.7 (normalization)$.

In order to calculate invariant yields, the transverse
momentum distribution of the measured $\overline{d}$s must also be understood
as well as possible. Because it
is not known which candidates under the mass peak are the true $\overline{d}$s,
the $p_t$ distribution is calculated for random selections of the candidates.
For each selection, the acceptance-corrected count is determined by
acceptance correcting every candidate at its measured $y$ and $p_t$.
Millions of random selections are made and a distribution of
acceptance-corrected counts is found with a most probable value and
widths which indicate the systematic errors of the method.
Between the rapidities of
1.8 and 2.2, the candidates have an acceptance-weighted $<$$p_t$$>$ of
0.35~GeV/c, and have an upper limit of $p_t = 1$~GeV/c.
Correcting for all efficiencies results in a $\overline{d}$ invariant yield of
$3.5 \pm 1.5 (stat.) ^{+0.9} _{-0.5} (sys.) \times 10^{-8}$ GeV$^{-2}c^{2}$.
%The systematic errors are from acceptance correction only,
%and do not include errors of $\sim$10\% from normalization to the beam.

In the rapidity range between 1.4 and 1.8,
the candidates have an acceptance-weighted $<$$p_t$$>$ of
0.26~GeV/c, and have an upper limit of $p_t = 0.5$~GeV/c.
All possible selections of the candidates are tried, determining a yield of
$3.7 \pm 2.7 (stat.) ^{+1.4} _{-1.5} (sys.) \times 10^{-8}$ GeV$^{-2}c^{2}$.

\begin{figure}
\epsffile{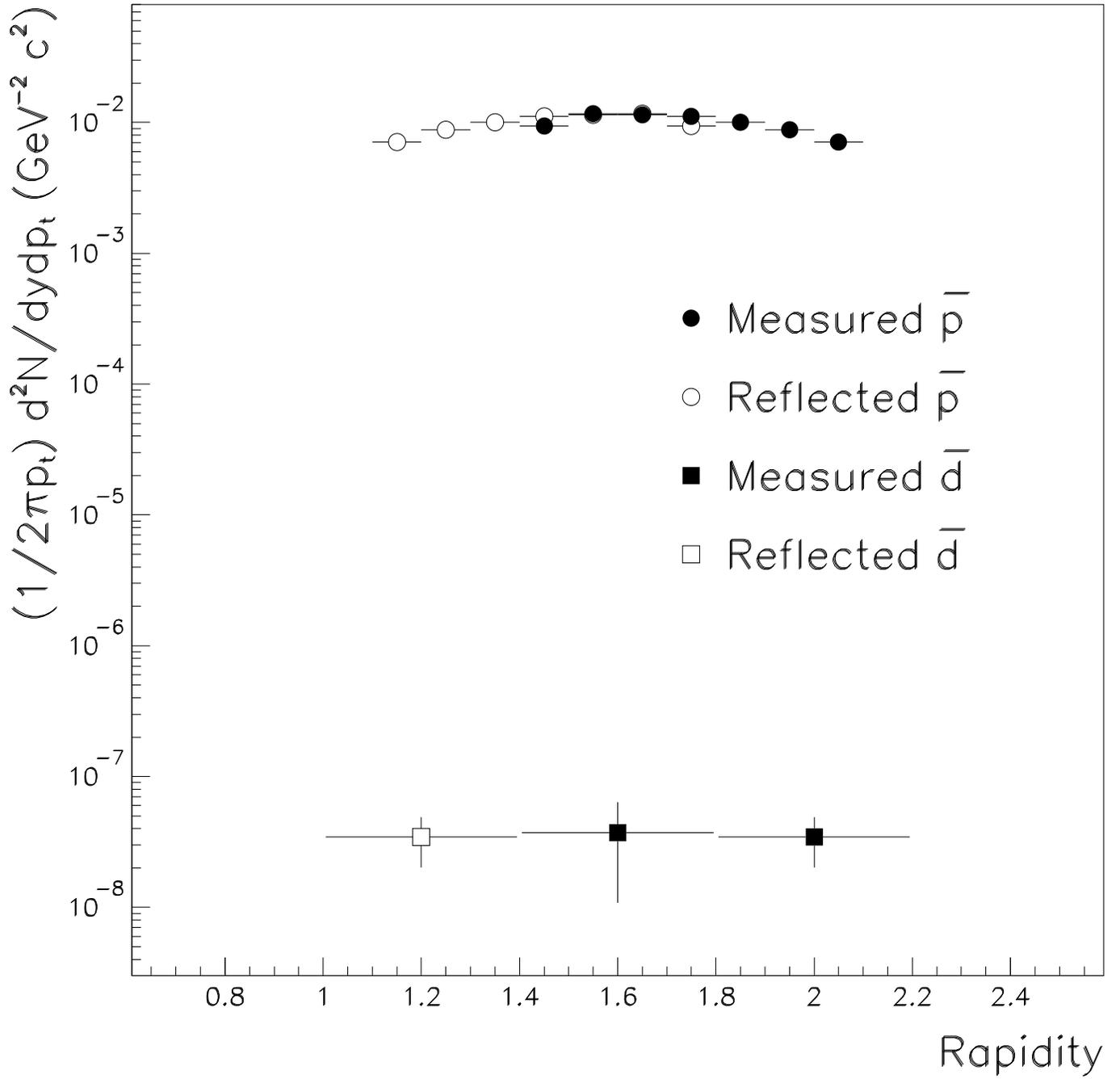}
\caption{Antimatter invariant yields measured by E864 in 10\% central
heavy ion collisions (statistical errors only).}
\label{fi:yield}
\end{figure}

E864 has also measured $\overline{p}$ yields in 10\% central Au~+~Pb collisions
at 11.5~GeV/c per nucleon~\cite{jl}. These yields are shown along with the
new $\overline{d}$ measurements in Figure~\ref{fi:yield} reflected about
mid-rapidity. The new $\overline{d}$
yields are in agreement with upper limits previously published by E864
taken from a smaller data sample~\cite{jl}.

The measured $\overline{p}$ and $\overline{d}$ yields allow us to calculate the
coalescence factor $\overline{B_2}$ using Equation~\ref{eq}. The E864
$\overline{p}$ measurement must be corrected for contributions from
antihyperon decays as these do not participate in the coalescence process.
Here, we will use the most probable value for
$\overline{Y}/\overline{p} = 3.5$, and use the 98\% confidence limit
of $\overline{Y}/\overline{p} > 2.3$ to define the systematic error of the
correction~\cite{jl}. This is a significant correction made from
an indirect measurement of $\overline{Y}/\overline{p}$ which
attributes the entire difference in $\overline{p}$ yields between two
experiments to their acceptance for antihyperon decay contributions,
introducing sizable uncertainties into our calculation of $\overline{B_2}$.
Additionally, as coalescence is a process affecting
co-moving nucleons, yields must be taken at the same $<$$p_t$$>$$/$$A$.
In the mid-rapidity bin ($1.4 < y < 1.8$), this means using the $\overline{p}$
yield at $p_t = 0.13$~GeV/c,
for which E864's $\overline{p}$ measurements are valid.
The final value of $\overline{B_2}$ comes out as
$4.1 \pm 2.9 (stat.) ^{+2.3} _{-2.4} (sys.) \times 10^{-3}$ GeV$^{2}c^{-3}$,
where the systematic error includes contributions from the $\overline{d}$
acceptance correction and the
antihyperon-feeddown correction to the $\overline{p}$s.

\begin{figure}
\epsffile{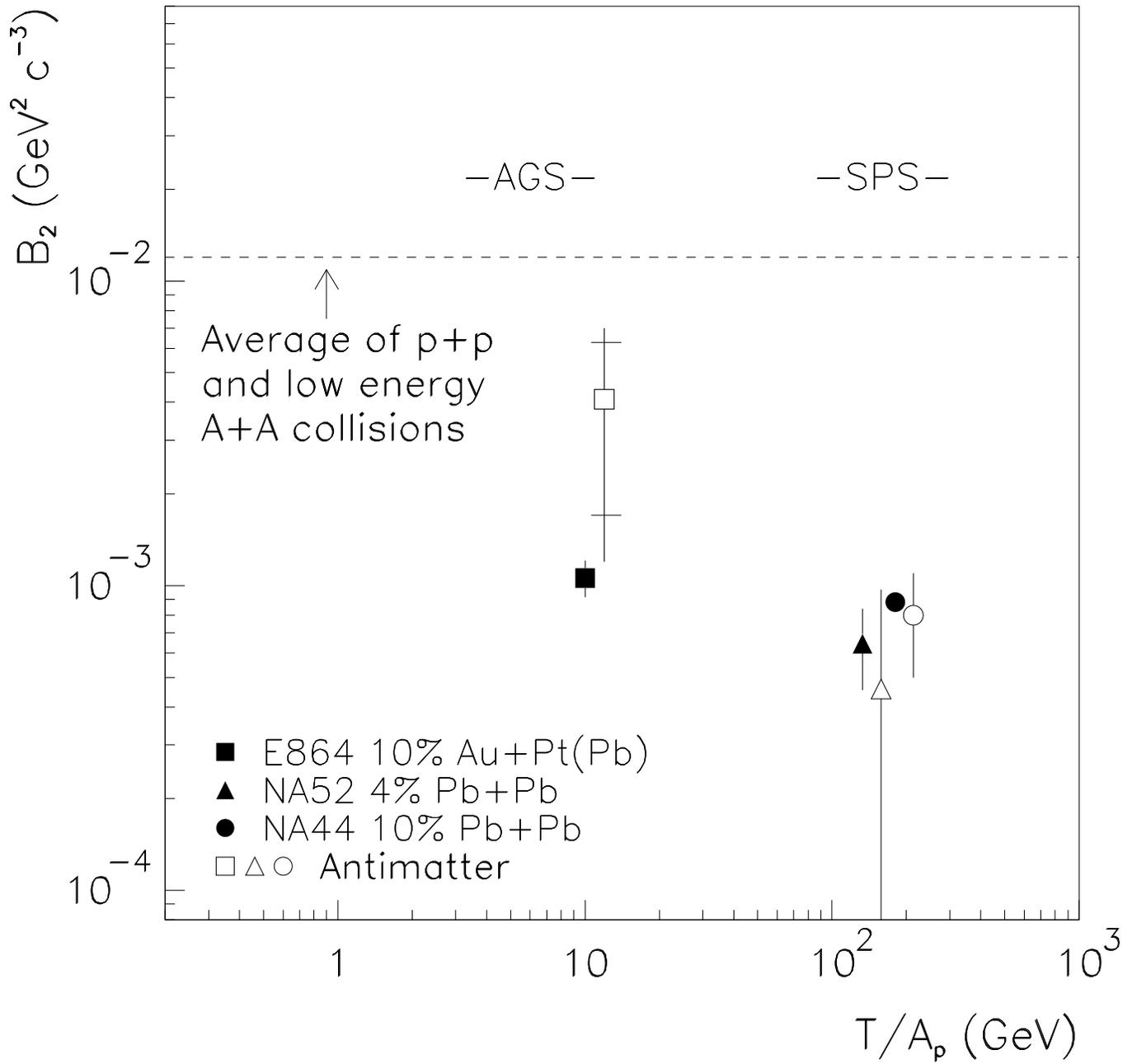}
\caption{Coalescence parameters for deuterons and antideuterons
at the AGS and SPS~\protect\cite{ng,sk,ib}.
Systematic errors of the E864 $\overline{B_2}$ measurement
are shown as horizontal bars.}
\label{fi:b2}
\end{figure}

E864 has also measured $B_2$ in the range $0.1 < p_t/A < 0.2$ GeV/c as
$1.06 \pm 0.15 \times 10^{-3}$ GeV$^{2}c^{-3}$~\cite{ng}.
While E864's $\overline{B_2}$ measurement is above its
$B_2$ measurement, the two are within statistical and
systematic errors of each other.
These values are shown in Figure~\ref{fi:b2} along with those
from other experiments studying collisions with
very large numbers of participant nucleons (${\sim}400$).
Along with the SPS results, the evidence
at this point suggests that there may be no difference between coalescence
of matter and antimatter in central heavy ion collisions.
This may be an indication that the freeze-out hypersurface for
antideuterons is not substantially modified by antimatter annihilation
from that of deuterons, and is consistent with predictions based on
a thermalized source~\cite{heinz2}. Predictions of
$\overline{B_2}$ supression from $B_2$~\cite{heinz,sm,surf}~are dependent
on such modifications and diminish in this scenario.

\section{SUMMARY AND ACKNOWLEDGEMENTS}

E864 has measured an antideuteron signal in central heavy ion collisions
at the AGS, where coalescence is likely to be the dominant
method for production. The measured invariant yields are
$3.5 \pm 1.5 (stat.) ^{+0.9} _{-0.5} (sys.) \times 10^{-8}$ GeV$^{-2}c^{2}$
($1.8 < y < 2.2$, $<$$p_t$$> = 0.35$~GeV/c) and
$3.7 \pm 2.7 (stat.) ^{+1.4} _{-1.5} (sys.) \times 10^{-8}$ GeV$^{-2}c^{2}$
($1.4 < y < 1.8$, $<$$p_t$$> = 0.26$~GeV/c).
The measured coalescence parameters from E864
for matter ($B_2 = 1.06 \pm 0.15 \times 10^{-3}$ GeV$^{2}c^{-3}$)
and antimatter ($\overline{B_2} =
4.1 \pm 2.9 (stat.) ^{+2.3} _{-2.4} (sys.) \times 10^{-3}$ GeV$^{2}c^{-3}$)
at mid-rapidity
are within statistical and systematic errors of each other.

We gratefully acknowledge the excellent support of the AGS staff. This work
was supported in part by grants from the U.S. Department of Energy's High 
Energy and Nuclear Physics Divisions, the U.S. National Science Foundation.

\small{
\begin{description}
\item[$\ast$]{Present address: Vanderbilt University, 
Nashville, Tennessee 37235 }
\item[$\dag$]{Present address: Istituto di Cosmo-Geofisica del CNR, Torino,
Italy / INFN Torino, Italy}
\item[$\ddag$]{Present address: Anderson Consulting, Hartford, CT}
\item[$\S$]{Present address: Univ. of Denver, Denver CO 80208}
\item[$\|$]{Deceased.}
\item[$\P$]{Present address: Cambridge Systematics, Cambridge, MA 02139}
\item[$\ast\ast$]{Present address: McKinsey \& Co., New York, NY 10022}
\item[$\dag\dag$]{Present address: Department of Radiation Oncology,
Medical College of Virginia, Richmond, VA 23298}
\item[$\ddag\ddag$]{Present address: University of Tennessee,
Knoxville, TN 37996}
\item[$\S\S$]{Present address: Institut de Physique Nucl\'{e}aire,
91406 Orsay Cedex, France}
\item[$\|\|$]{Present address: Institute for Defense Analysis,
Alexandria, VA 22311}
\item[$\P\P$]{Present address: MIT Lincoln Laboratory, 
Lexington, MA 02420-9185}
\end{description}
}

\clearpage

\clearpage

\clearpage

\clearpage

\end{document}